\documentclass{ws-p8-50x6-00}

\newcommand{ \rts }{$\sqrt{s_{_{\rm NN}}}$ }
\newcommand{ \pt }{$p_t$ }

\begin{document}
\title{Recent results from heavy ion collisions}
\author{Nu Xu}
\address{Lawrence Berkeley National Laboratory, Berkeley, CA 94720,
  USA \\ E-mail: nxu@lbl.gov}

\maketitle

\abstracts{Systematic trends of baryon transport, chemical
  freeze-out, and kinetic freeze-out in high energy nuclear collisions
  are presented. Further measurements of particles with heavy flavors
  are proposed in order to shed light on collision dynamics at parton
  level.}

\section{Introduction}

The purpose of the current heavy ion programs at CERN (Switzerland)
and Brookhaven National Laboratory (USA) is to probe strongly
interacting matter under extreme conditions, i.e.  at high densities
and temperatures. The central subject of these studies is the
transition from the quark-gluon plasma to hadronic matter.  In the
early phases of ultra-relativistic heavy ion collisions, when a hot and
dense region is formed in the center of the reaction, there is copious
production of up, down, and strange quarks.  Transverse expansion is
driven by multiple scattering among the incoming and produced
particles.  As the medium expands and cools, the quarks/gluons combine
to form the hadrons that are eventually observed.
  
In this paper, we will summarize the recent experimental results on
transverse momentum distributions and particle ratios. The related
physics issues are chemical equilibrium and collective expansion.
While the former has to do with inelastic collisions, the later is
dominated by the elastic cross section. We will focus at the
relatively low transverse momentum region, \pt $\le 2$ GeV/c, where
the bulk production occurs. Experimental results are from refs. [1-9]
and for results on high momentum transfer, global measurements, and
correlations readers are referred to ref. [10].

\section{Baryonic Physics: mid-rapidity $\bar{p}/p$ ratios and
  inclusive net-proton distributions}

All RHIC experiments have consistently measured the ratio of
$\bar{p}/p$ at
mid-rapidity\cite{starpbarp,phobospbarp,brahmspbarp,barish01}.  It is
found that at mid-rapidity, the ratio is not unity and a dramatic
increase in the ratio is observed from SPS to RHIC, meaning that the
system created at \rts = 130 GeV is not yet net-baryon
free\cite{starpbarp}.

One of the unsolved problems in high energy nuclear collisions is the
baryon transfer that occurs at the early stage of the
collision\cite{buzsa}. The later dynamic evolution of the system is
largely determined at this moment since the available energy is fixed
at this stage of the collision. On the theoretical side, a novel
mechanism, the non-perturbative gluon junction, was proposed to
address this problem\cite{dima96,vance98}. From the STAR measured
$\bar{p}/p$ ratios\cite{starpbarp} and the STAR preliminary results
of anti-proton yields\cite{starpbar}, we have extracted the
net-proton yields at mid-rapidity. The values are shown in Fig.\,1 as
a function of collision centrality. No hyperon decay corrections have
been applied as that requires the knowledge of the yields of hyperons.

\vspace{-1.250cm}

\begin{figure}[h]
\hspace{1.cm}
\epsfxsize=25pc 
\epsfbox{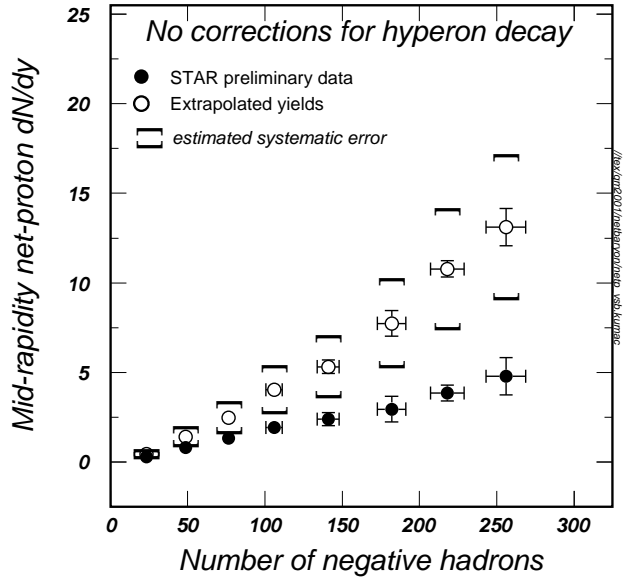}
\vspace{-1.5cm}
\caption{\it Inclusive net-protons yields vs. charge
  particle multiplicity. Open and filled symbols represent the yields
  within $p_t \le$ 1 GeV/c and the values extracted from the full
  transverse momentum range, respectively.}
\end{figure}

One can see that the number of net-protons increases as the number of
the negatively charged particles increases. This implies that more and
more protons (baryons) are transfered from the incoming nuclei to
mid-rapidity as the overlap of the two nuclei increases. More data are
needed to study the role of the junction
mechanism\cite{dima96,vance98} in heavy ion collisions. In this
respect, the rapidity distributions of the net-protons as a function
of collision centrality will be of crucial importance.


\section{Chemical Freeze-out: particle ratios}

Recently, much theoretical effort has been devoted to the analysis of
particle production within the framework of statistical
models\cite{pbm9596,becattini98,cleymans98}. These approaches are
applied to the results of both elementary collisions ($e+e^-$, $p+p$)
and heavy ion collisions ($Au+Au$ and $Pb+Pb$)\cite{pbm9596}.  Many
features of the data imply that a large degree of chemical
equilibration may be reached both at AGS and SPS energies. The three
most important results are: (i) at high energy collisions the chemical
freeze-out (inelastic collisions cease) occurs at about 160-180 MeV
and it is `universal' to both elementary and heavy ion collisions;
(ii) the kinetic freeze-out (elastic scatterings cease) occurs at a
lower temperature $\sim 120 - 140$ MeV; (iii) the compilation of
freeze-out parameters\cite{cleymans98} in heavy ion collisions in the
energy range from 1 - 200 A$\cdot$GeV shows that a constant energy per
particle $\langle E \rangle / \langle N \rangle \sim 1$ GeV can
reproduce the behavior in the temperature-potential ($T_{ch} -
\mu_{_B}$) plane\cite{cleymans98}.

\vspace{-1.5cm}

\begin{figure}[hb]
\hspace{1.cm}
\epsfxsize=25pc 
\epsfbox{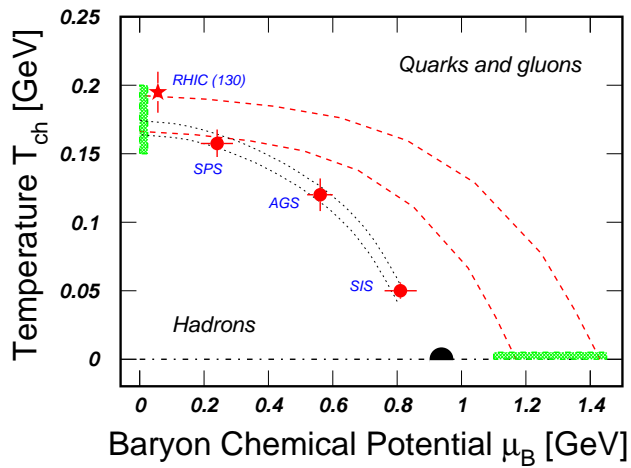}
\vspace{-1.5cm}
\caption{\it Phase plot $T_{ch}$ vs. $\mu_{_B}$.
  Dashed-line represents the boundary between interactions involving
  hadronic and partonic degrees of freedom.  Dotted-lines represent
  the results of [21].  The ground state of nuclei is shown as half
  circle.}
\end{figure}

The systematics for chemical parameters is shown in the phase plot in
Fig.\,2 where the dashed lines represent the boundary between
interactions involving hadronic and partonic degrees of freedom.  The
dotted lines represent the results of  ref. [21]. The lattice
prediction on $T_c$ ($\approx 160$ MeV) is shown at $\mu_B = 0$ and
the baryon density for neutron stars is indicated at $T_{ch} = 0$.

\section{Kinetic Freeze-out: transverse momentum distributions}

The measured transverse momentum distributions have been fitted by the
exponential function $f = A \cdot exp(-m_t/T),$ where $T$ is the slope
parameter and $A$ is the normalization constant.  The magnitude of the
slope parameter provides information on temperature (random motion in
local rest frame) and collective transverse flow.  Fig.\,3(a) shows
the measured particle slope parameters from Pb+Pb central collisions
at SPS (\rts = 8.8 GeV and \rts = 17.2 GeV)
energies\cite{harry01,blume01,na44_coll_exp,tsukuba}. Recent results
are:

\begin{itemize}
\item Particle distributions from 40 A$\cdot$ GeV (\rts = 8.8 GeV)
  collisions were reported by NA45 and NA49\cite{harry01,blume01}.
  The slope parameters are similar to the ones observed at 158A$\cdot$
  GeV, {\it i.e.}, they follow the established systematic
  trend\cite{nxuqm96};
  
\item The systematic of the transverse momentum
  distributions for $J / \psi$ was reported by the
  NA50 experiment\cite{bordalo01,na50plb}. It is interesting to note
  that the slope parameter of $J / \psi$ is similar to those for
  $\phi$ and $\Omega$;
  
\item For central collisions the $\phi$ slope parameter of NA49 is
  about 300 MeV\cite{friese01} whereas that of the NA50 is about 240
  MeV.  NA49 and NA50 reconstructed $\phi$ mesons via $K^+K^-$ and
  $\mu^+\mu^-$ channels, respectively.  As discussed in
  refs. [25,26], part of the difference may be caused by
  the final state interaction of the decay kaons. In fact, if one
  studies the centrality dependence of the $\phi$ slope parameter, one
  finds that at peripheral collisions, both experimental results agree
  with each other and the value is close to that from $p+p$
  collisions\cite{friese01,bordalo01}.
\end{itemize}

\begin{figure}[h]
\hspace{-0.72cm}
\epsfxsize=30pc 
\epsfbox{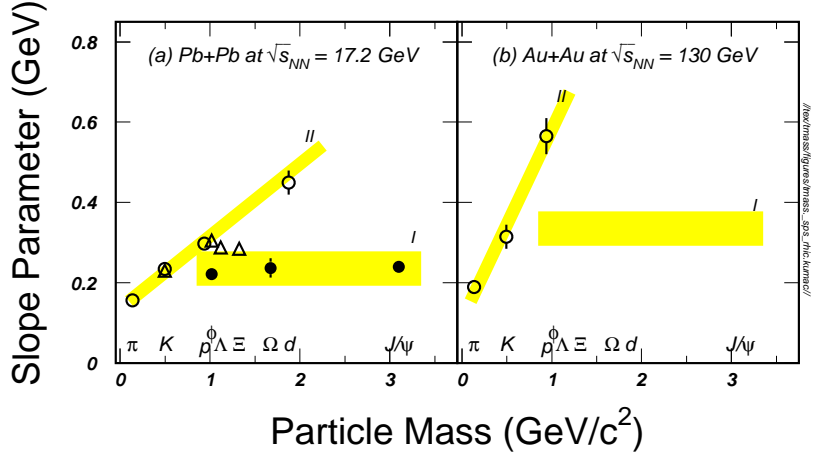}
\vspace{-1.5cm}
\caption{\it Slope parameters as a function of particle
  mass for (a) Pb+Pb central collisions at SPS (\rts = 17.2 GeV) and
  Au+Au central collisions at (b) RHIC (\rts = 130 GeV).  Weakly and
  strongly interacting limits are indicated by $I$ and $II$,
  respectively.}
\end{figure}


Preliminary results of the slope parameters from Au+Au central
collisions at RHIC (\rts = 130 GeV)\cite{harris01} are shown
in Fig.\,3(b).  It is obvious that the mass dependence of the slope
parameter is stronger than that from collisions at SPS energies.

As one can see from Fig.\,3(a), the slope parameters appear to fall
into two groups: group (I) is flat as a function of particle mass,
whereas the slopes in group (II) increase strongly with particle mass.
At the RHIC energy, the slope parameter systematic of $\pi, K,$ and
$p$ shows an even stronger dependence on the particle mass.  The
strong energy dependence of the slope parameter might be the result of
the larger pressure gradient at the RHIC energy. With a set of
reasonable initial/freeze-out conditions and equation of state, the
stronger transverse expansion at the RHIC energy was, in deed,
predicted by hydrodynamic calculations\cite{deark01,passplb01}. In
addition, the results are consistent with the large value of the event
anisotropy parameter $v_2$ at the RHIC
energy\cite{starflow,snelling01}.

Within the hadronic gas, the interaction cross section for particles
like $\phi, \Omega,$ and $J / \psi$ are smaller than that of $\pi, K,$
and $p$\cite{hsx98}.  Therefore the interactions between them and the
rest of the system are weak, leading to the flat band behavior in
Fig.\,3(a).  On the other hand, the slope parameter of these weakly
interacting particles may reflect some characteristics of the system
at hadronization. Then it should be sensitive to the strength of the
color field\cite{schwinger,bleicher001}. Under this assumption, the
fact that the weakly interacting particles show a flat slope parameter
as a function of their mass would indicate that the $flow$ develops at
a later stage of the collision. Should the collective flow develop at
the partonic level, one would expect to see a mass dependence of the
slope parameters for all particles\cite{thews01}.

Figure 4 shows the bombarding energy dependence of the freeze-out
temperature $T_{fo}$ and the average collective velocity $\langle
\beta_t \rangle$. These parameters were extracted by hydrodynamic
motivated model fitting\cite{heinz93}. It is interesting to observe
that both $T_{fo}$ and $\langle \beta_t \rangle$ saturate at a beam
energy of about 10 A$\cdot$GeV. The saturation temperature is about
100 - 120 MeV, very close to the mass of the lightest
meson\cite{pomeranchuk}.  The steep rise of the $T_{fo}$ and $\langle
\beta_t \rangle$ up to about 5 A$\cdot$GeV incident energy indicates
that at low energy collisions the thermal energy essentially goes into
kinetic degrees of freedom.  The saturation at $\sim$10 A$\cdot$GeV
shows that particle generation becomes important.  As proposed
in\cite{pomeranchuk,heg}, for a pure hadronic scenario there may be a
limiting temperature $T_c \approx 140$ MeV in high-energy collisions,
although the underlying physics for both, the transition from partonic
to hadronic degrees of freedom and the transition from interacting
hadrons to free-streaming is not clear at the moment.  By coupling the
limiting temperature idea to a hydrodynamic model calculation,
St\"ocker {\it et al.}  successfully predicted\cite{stocker} the
energy dependence of the freeze-out temperature.

\vspace{-1.05cm}

\begin{figure}[h]
\hspace{-0.75cm}
\epsfxsize=27.5pc 
\epsfbox{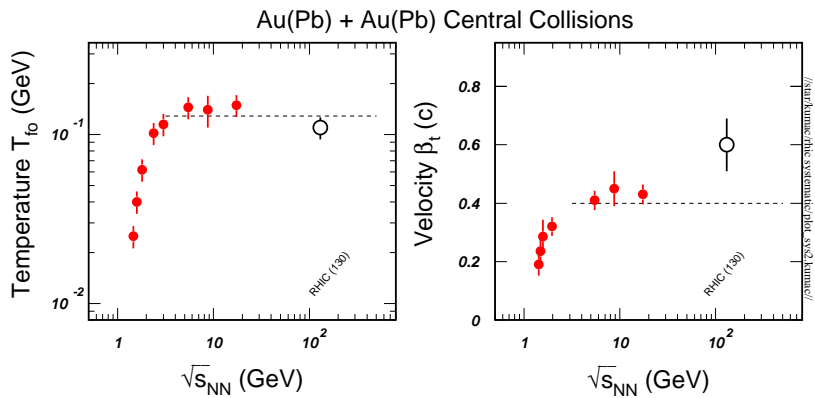}
\vspace{-0.75cm}
\caption{\it Systematics of kinetic freeze-out
  temperature parameter $T_{fo}$ and average collective transverse
  flow velocity $\langle \beta_t \rangle$ as a function of beam
  energy.  At \rts $\approx$ 5 GeV, both values of temperature and
  velocity parameters seem to saturate.  However, the velocity
  parameter extracted from central collisions at the RHIC energy is
  higher than values from collisions at lower beam energies.
  \label{fig:fig7}}
\end{figure}

At the RHIC energy, the collective velocity parameter seems to be
larger than that from collisions at AGS/SPS energies. This can already
be seen in Fig.\,4 where the increase from pion to proton at \rts =
130 GeV (Fig.\,3(b)) is much faster than at \rts = 17 GeV
(Fig.\,3(a)). Is this a manifestation of van Hove's\cite{vH83}
picture? On the other hand, compared to results from lower energy
collisions, the temperature parameters seem to be lower. Is this the
consequence of hydrodynamic expansion from a higher
initial density fireball? An energy scan between \rts = 20 - 130 GeV
will be extremely important in order to study this evolution in more
detail.

\section{Summary}

In summary, the most interesting results from collisions at RHIC are
that the system is indeed approaching net-baryon free and the
transverse expansion is found to be stronger than that from collisions
at AGS/SPS energies.

In order to understand the trend of the collective velocity, an energy
scan between \rts = 20 - 200 GeV, is important. In addition,
systematic studies on the anisotropy parameter $v_2$ and the
transverse momentum distributions of $\phi, \Omega,$ and $J / \psi$
are necessary as they will help in determining whether the
collectivity is developed at the partonic stage.

 -----
 
 We are grateful for many enlightening discussions with Drs. P.
 Braun-Munzinger, M. Gyulassy, H.G.  Ritter, K.  Schweda, E.V.
 Shuryak, F.Q.  Wang, and X.N.  Wang.  This work has been supported by
 the U.S.  Department of Energy under Contract No. DE-AC03-76SF00098.


\end{document}